\documentclass[11pt]{article}
\usepackage[cp866]{inputenc}

\makeatletter

\renewcommand{\@oddhead}{\hfil Astrophysics of Wormholes}

\input epsf

\begin{document}

December 24, 2006

\begin{center}{\bf  {\Large
Astrophysics of Wormholes}\\
\quad \\
N.S. Kardashev${}^1$, I.D. Novikov${}^{1,2}$, and A.A.
Shatskiy${}^1$}
\end{center}
{\bf ${}^1$} Astro Space Center, Lebedev Physical Institute, Russian Academy of Sciences,\\
Profsoyuznaya ul., 84/32, Moscow, 117997, RUSSIA. \\
{\bf ${}^2$} Niels Bohr Institute, Blegdamsvej 17, DK-2100,
Copenhagen Denmark.
\\

{\bf\large ABSTRACT}\\
We consider the hypothesis that some active galactic nuclei and
other compact astrophysical objects may be current or former
entrances to wormholes. A broad mass spectrum for astrophysical
wormholes is possible. We consider various new models of the
static wormholes including wormholes maintained mainly by an
electromagnetic field. We also discuss observational effects of a
single entrance to wormhole and  a model for a binary
astrophysical system formed by the entrances of wormholes with
magnetic fields and consider its possible manifestation.

\section{INTRODUCTION}
\label{s1} 
\begin{figure}[t]
\centering \epsfbox[100 250 550 550]{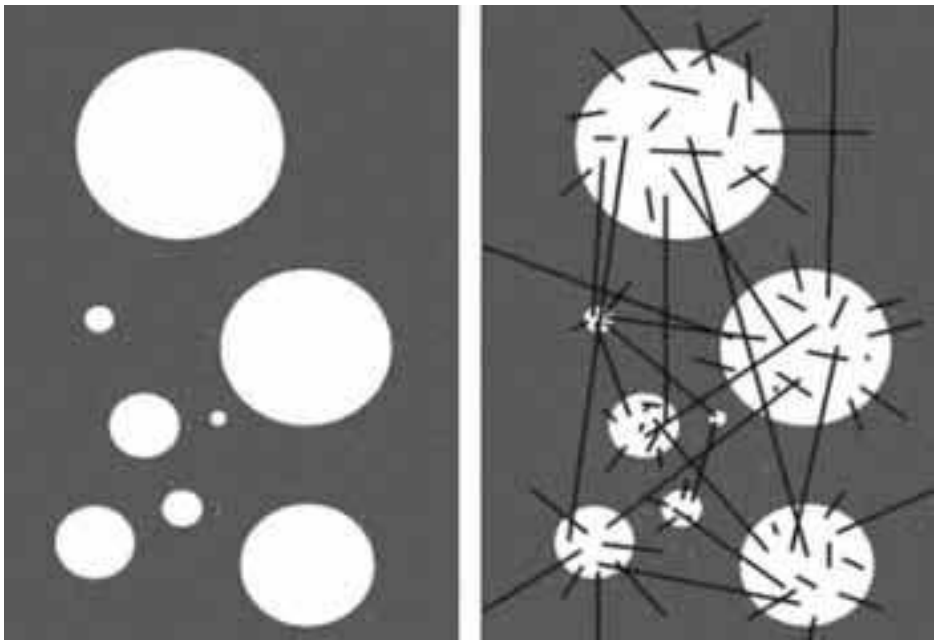} \caption{Model of
chaotic inflation in a multi-element Universe without (left) and
with (right) wormholes.} \label{R1}
\end{figure}

The purpose of this paper is to consider a possibility
that some astrophysical objects may be current or former entrances
to wormholes (WHs). These wormholes may be relic of the inflation
epoch of the evolution of the Universe \cite{1} - \cite{9} and
fig.~\ref{R1}.

It follows from WH models that their existence requires matter
with a peculiar equation of state \cite{10}-\cite{46}. This
equation must be anisotropic, and ${w_\parallel =p_\parallel
/\varepsilon}$ must be smaller than ${-1}$, as in the case of
phantom matter (${p_\parallel}$ is the total pressure along the
tunnel of the WH, and ${\varepsilon}$ is the total energy density
for all components of the matter in the tunnel of the WH). The
existence of such matter remains hypothetical \cite{14}. For
definiteness, we will use the term "phantom energy"\, for an
isotropic equation of state ${p/\varepsilon < -1}$, and the term
"phantom matter"\, for an anisotropic equation of state. The units
are selected so that ${c = 1}$ and ${G = 1}$ (with exceptions for
 final relations).

In this paper we consider a model in which the main component of a
wormhole having all the necessary properties is a strong magnetic
field that penetrates the WH, while phantom matter and phantom
energy are required only in small amounts. We also consider a
model based on phantom energy, where the equation of state is
close to that for a vacuum ${(p/\varepsilon =-1)}$, with some
added energy density of the magnetic field.

We do not consider here the dynamics of wormholes. Thus all
conditions for the relations between the components of the
stress-energy tensor correspond to the static state but not
dynamics.

In particular we do not consider the problem of stability of the
models of wormholes. It is necessary to emphasize that there are
stable models of wormholes (see for example \cite{Armen}).

We consider the corresponding model of WHs, and their properties
in Appendixes. For an external observer, the entrance to the
wormhole appears to be a magnetic monopole with a macroscopic
size. The accretion of ordinary matter onto the entrance to the
wormhole may resulting the formation of a black hole with a radial
magnetic field. We consider the possibility that some active
galactic nuclei and some of Galactic objects may be current or
former entranced to magnetic wormholes. We consider the possible
existence of a broad mass spectrum for wormholes, from several
billion solar masses to masses of the order of 2 kg. The Hawking
effect (evaporation) does not operate in such objects due to the
absence of a horizon, making it possible for them to be retained
over cosmological time intervals, even if their masses are smaller
than ${10^{15}g.}$ We also discuss a model for a binary system
formed by the entrances of tunnels with magnetic fields, which
could be sources of non-thermal radiation and $\gamma$-ray bursts.

There is a hypothesis that the primordial wormholes probably
exists in the initial state of the expanding Universe
\cite{6},\cite{7} and can connect different regions in our
Universe and other Universes in the model of Multiverse. It is
possible that primordial WHs are preserved after inflation. In
this case the search for astrophysical wormholes is a unique
possibility to study the Multiverse.

Some aspect of the problem you can find in \cite{our}.

\section{WORMHOLES AND THEIR REMNANTS IN THE UNIVERSE}
\label{s2} As we emphasized in Section~\ref{s1} our hypothesis
suggests that it may be possible existence in the modern Universe
entrances to the wormholes with rather strong monopole magnetic
fields. Explicit models of them are in Appendix \ref{a1}.
Astrophysical accretion of ordinary matter on them may probably
converts them into black holes (BHs) with strong monopole magnetic
field (wormhole remnants). The monopole magnetic field of the WH
remnants differs from ordinary BHs. The last ones nether have
strong monopole magnetic field. Probably some of Galactic and
extragalactic objects are such WH/BHs. Detecting these WH/BHs is a
great challenge.

\begin{figure}[t]
\centering \epsfbox[20 300 500 600]{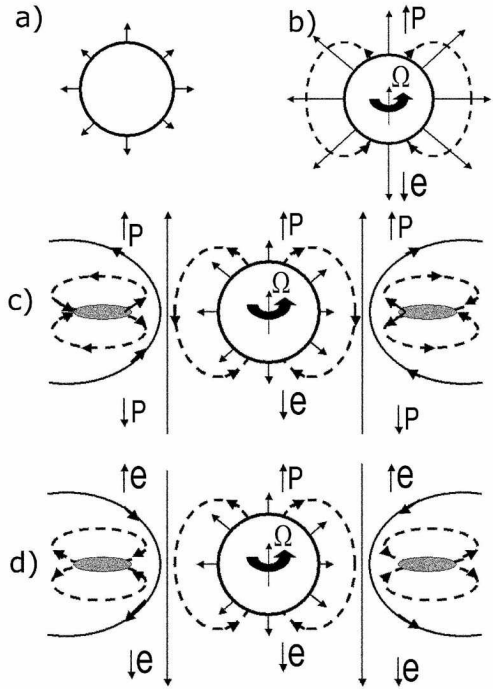} \caption{A schematic
toy model of the WH/BH: (a) WH or BH with a radial magnetic field.
(b) Same as (a), with rotation and a dipolar electric field. (c)
Same as (b), with a rotating accretion disk with its dipole
magnetic field and quadrupolar electric field. (d) Same as (c),
with dipole magnetic and quadrupole electric fields of the
opposite direction. The arrows (p) and (e) indicate the directions
for proton and electron jets, respectively.} \label{R3}
\end{figure}

Figure \ref{R3} presents WH/BH models allowing for the possibility
that they possess an evolved accretion disk with its own magnetic
field (probably dipolar and substantially weaker than the field of
the WH/BH themselves). This is a schematic toy model in the vacuum
approximation. The presence of a radial magnetic field can be
revealed by the law for the increase in the field strength
${(H\propto r^{-2})}$ and the presence of the same sign of the
field on all sides. The rotation of the monopole excites a dipolar
electric field, which can provide acceleration for relativistic
particles. Note that the dipole electric field (in contrast to the
quadruple field for a disk) accelerates electrons towards one of
the poles (and protons and positrons towards the other; see
Appendix \ref{a3}). Of course such motion of electrons is possible
only if there are currents providing the electro neutrality of the
system. Something like this may explain the origination of
one-sided jets in some sources (for example, the quasar 3C273)
\cite{30}.

Actually the general picture should be much more complicated.

First of all the rarefied plasma form a complex magnetosphere
around WH/BH (see \cite{41}). In addition the presence of the
accretion disk complicates the picture. In a toy model the
quadrupolar electric field generates two-sided electron or
proton/positron jets (depending on the sign of the quadrupole). As
a result, the structure of the jets may be different:

1) electrons from the magnetic WH/BH are ejected from one pole,
while protons (positrons) are ejected from the other,

2) electrons from the WH/BH and the accretion disk are ejected
from one pole, while protons (positrons) from the WH/BH and
electrons from the accretion disk are ejected from the other,

3) electrons from the WH/BH and protons (positrons) from the
accretion disk are ejected from one pole, while protons
(positrons) from the WH/BH and accretion disk are ejected from the
other. The interactions between the electromagnetic fields of the
WH/BH and accretion disk are likely to be very strong.

We have to remember that even in the toy model such jets are
possible only under the conditions of existence of the electric
currents providing the electro neutrality of the system and
inducted electric field conserves.

The difference between a WH entrance and a BH may be revealed from
observations indicating the absence of a horizon --- a source of
light falling into aWH should be observed continuously, but with a
variable red, and even blue, shift. However, in this case, we must
assume the tunnel is transparent. A blue shift can appear if the
mass of the further WH entrance (relative to the observer) exceeds
that of the closer entrance. If the tunnel is transparent and has
accretion disks at both entrances, the redshift of the spectra of
these disks will also be different. Thus, two different redshifts
from a single source related to aWH could be distinguished.

An observed WH image may display some internal structure, whose
angular size could be substantially smaller than that specified by
the gravitational diameter.

In this connection, observations of the gravitationally lensed
quasar Q0957+561 with the redshift ${z=1.4141}$~\cite{31} are of
great interest. The observation show that in this active galaxy
exist the central compact object with mass ${3.6\cdot
10^{9}M_\odot}$, strong magnetic moment and without an event
horizon. A specific property ofWHs, which display strong
relativistic effects and also the absence of a horizon, is the
possibility of periodic oscillations of a test mass relative to
the throat (See TABLE 1 and Appendix~\ref{a1}). The redshift will
also vary periodically. If the structure of the WH is close to the
formation of the horizon, these shifts, the period, and the flux
variations can be very large. We point out in this connection the
quasi-periodic flux variations observed for so-called IDV sources,
such as the BLLac object 0716+714~\cite{32}.

\begin{figure}[t]
\centering \epsfbox[200 250 320 550]{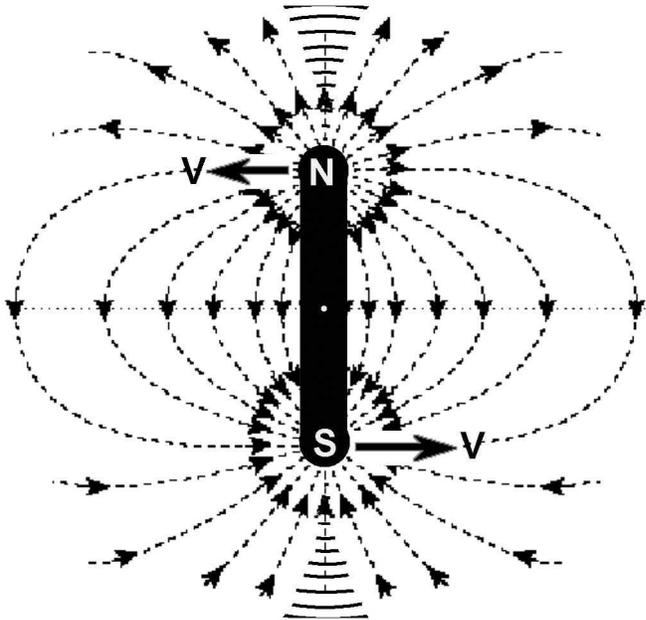} \caption{Schematic
model of a binary system of two magnetic WHs/entrances.}
\label{R4}
\end{figure}

When sources move along circular orbits around a WH entrance (the
laws of motion see in~\cite{our} and Appendix~\ref{a4}), the flux
and redshift of radiation of a compact source will also vary
periodically.

Ultimately, an external observer may be able to detect some
radiation at the gyrofrequencies and events related to the
creation of ${e^\pm}$ and ${\mu^\pm}$ pairs (where $\mu$
designated a magnetic monopole).

Let us also consider the possibility of binary objects formed by
two entrances to the same WH that interact gravitationally and
electromagnetically and rotate about their common center of
gravity. If the two entrances in the model with a magnetic field
represent magnetic monopoles of different signs, then the binary
forms a rotating magnetic dipole. Such binaries with
closely-spaced entrances are also very likely to originate in the
primordial scalar field, and can be preserved until the present
(Fig.~\ref{R4}).

Let us denote: $d$ --- is the distance between the WH entrances,
$T$ --- is the period of the circular orbit, and ${M_\infty}$ ---
is the asymptotic mass at infinity of each entrance.

As it is shown in Appendix~\ref{a1} in our model the magnetic
field ${H_0}$ at the throat of the WH is related with
${M_\infty}$:
\begin{equation}
H_0\approx \left( c^4/G^{3/2}\right)\, M_\infty^{-1}
\label{n1}\end{equation} Then for the circular orbit
(see~\cite{our}):
\begin{equation}
V=\sqrt{{GM_\infty\over d}}\, ,\quad T=\pi\sqrt{{d^3\over
GM_\infty}} \, . \label{n2}\end{equation} These relations are
derived in the Newtonian approximation and for motion under the
action of the gravitational and magnetic fields.

The intensity of magnetic-dipole ${I_{mag}}$ and gravitational
${I_{grav}}$ radiation is correspondingly~\cite{33}:
\begin{equation}
I_{mag}={2^5 G^3 M_\infty^4 \over 3 c^3 d^4} \, ,
\label{n3}\end{equation}
\begin{equation}
I_{grav}={2^9 G^4 M_\infty^5 \over 5\cdot c^5 d^5} \, .
\label{n4}\end{equation} Gravitational radiation dominates only at
very small distances ${d<(3/5)\cdot 2^{4}\cdot(GM_\infty /c^2)\le
9.6 r_0,}$ (see appendix~\ref{a1}) so that the losses of the
energy and evolution of the system are determined by the
magnetic-dipole radiation~(\ref{n3}).

The characteristic time of the evolution of the system due to the
radiation is ${t_{em}=|{\cal E}_{total}|/I_{mag},}$ where
\begin{equation}
{\cal E}_{total}=-M_\infty V^2 \label{n4-2}\end{equation} Hence
the system parameters preserved over time ${t_{em}}$ are no
smaller than
\begin{equation}
d^3_{em}={2^5 G^2 M_\infty^2 t_{em}\over 3 c^3}\, ,\quad
T_{em}=\pi\sqrt{{2^5 G M_\infty t_{em}\over 3 c^3}} \, .
\label{n5}\end{equation} A rotating system of macroscopic magnetic
dipole induces an electric field, which accelerate electrons and
emit electromagnetic radiation along the axis of the magnetic
dipole (similarly to pulsars) or even create of electron-positron
pairs. Corresponding critical electric field is ${{\cal
E}_{cr}=4\cdot 10^{13}.}$ Now if we specify ${t_{em}}$ equal
${t_{_U}}$ the age of the Universe ${t_{_U}\approx 13\cdot 10^9}$
years, we obtain the following characteristic of the system with
electric field ${{\cal E}_{cr}}$ and ${t_{em}=t_{_U}}$:
\begin{equation}
M_\infty =3.2\cdot 10^{35}g\approx 160\, M_\odot\, ,\quad
d_{em}=4.2\cdot 10^{14}cm \, ,\quad T_{em}=1.9\cdot 10^8 s\approx
6\, yrs \, ,\quad V_{em}\approx 70\, km/s \, .
\label{n6}\end{equation} All binary systems with smaller masses
will generate ${e^\pm}$, forming two-sided jets of relativistic
particles in the orbital plane, lose energy, and collapse.

TABLE 1 presents the parameters of magnetic WHs with various
masses (notations see in appendix~\ref{a1}) and estimates for the
intensity of magnetic-dipole radiation ${I_{mag}}$, the component
separation ${d_{em}}$, the rotational period ${T_{em}}$ and
velocity ${V_{em}}$ for systems preserved to the present epoch.

In this toy model binaries can be detected as strictly periodical
nonthermal sources (similar to pulsars) with two jets of
relativistic ${e^\pm}$ in the equatorial plane and a cloud of
annihilating low-energy particles ${e^\pm}$.

The approach of the two entrances of the same magnetic WHs ends
with its transformation into a BH with a mass of ${\sim M_\infty}$
without a magnetic field, with the radiation of the external
magnetic field ${{\cal E}_{mag,\, total}\approx M_\infty c^2}$,
similar to the collapse of a magnetized body \cite{34}, \cite{35}.
The electromagnetic impulse accelerates the created ${e^\pm}$
pairs, and may represent the basic mechanism for some type of
observable $\gamma$-ray bursts. The approach of two closely spaced
different WH entrances should be somewhat different: after they
merge, probably one entrance or a BH is formed, with its magnetic
flux equal to the sum of the original magnetic fluxes, taking into
account their signs and the radiation of the difference in the
flux. Thus, various merging processes could result in the
formation of more massive objects, within which the entrances and
magnetic fluxes from many tunnels are combined.

\section{WORMHOLES AND THE MULTI-UNIVERSE}
\label{s4} As we mentioned in section~\ref{s1}, there is a
hypothesis that a wormholes can connect different Universes in the
model of multi-universe (see fig.~\ref{R1}).

In this section briefly consider various aspects of such a
possibility. Let us consider the possible observational
differences of the two cases:

(a) both entrances of the wormhole are in our Universe  and

(b) one entrance is in our Universe and another entrance is in another Universe.\\
In case (a) both entrances probably were formed in the very early
quantum period of the existence of our Universe. It means that
both of them were formed practically at the same moment of time of
an external observer which is practically at rest with respect to
the entrances. If during the subsequent evolution of the Universe
there was not great relative velocity of the motion of the
entrances and if there was not great difference in gravitational
potential in their surrounding then in our epoch, time near each
entrance corresponds to the same cosmological time of our epoch.

Hence an observer looking through the throat of the wormhole sees
our Universe at present epoch (but another regions of it). The
radiation from the wormhole would be provided mainly by the
processes near both entrances rather then the cosmic microwave
background which is quite weak.

In case (b) situation may be quite different. In principle the
physics in our and other universe can be absolutely different and
in principle we can observe it. But even if we suppose that
another Universe is similar to our one there is not probably any
correlation in times in these universes. Thus in this case we can
see through the wormhole any epoch in the evolution of the Hot
Universe: (1) very early one, or (2) close to our epoch, or (3)
very late epoch.

In case (1) we would see the epoch which corresponds to very hot
background radiation. The flux of the radiation through the
wormhole into our Universe would rapid decrease the mass and the
size of the entrance in our Universe and would lead to rapid
conversion of it into (a) black hole or into a naked singularity.

The dynamics of space-time of a wormhole in the case of a strong
flux of matter through it needs a special investigation and we
will discuss it in a separate paper.

Here we want to mention that such an instability of a wormhole
with respect to flux of energy brought it may be very important
also in the case (a) at the very early epoch if there is not a
good balance of the contrary fluxes of radiation in both
entrances.

Case (2) would be similar to the case (a) from the point of view
of observations.

In case (3) we would see a very cold Universe without any
significant radiation from it.

It is worthy of consideration the following problem. The entrance,
in which the flux of radiation is falling down, would increase in
mass and in size. As we mentioned above the opposite entrance
would shrink down and eventually probably would disappear.

The question arises: what does an observer near first entrance see
when the second entrance disappears? This question is a part of
the problem about the evolution of the wormhole in the case of a
strong flux of energy through it and will be considered in a
separate paper.

\section{CONCLUSION}
\label{s3} In this paper we propose to analyze the possibility of
identifying a new type of primordial cosmological objects --- WHs
entrances, or BHs formed from WHs --- among known Galactic and
extragalactic objects that are usually identified with neutron
stars or stellar-mass and super massive BHs. The implied presence
of a strong radial magnetic field (the "hedgehog"\, model
\cite{35}, \cite{36}, \cite{37}) is very important, and can,
together with rotation, bring about the generation of (one- and
two-sided) relativistic jets. WH models suggest the specific
properties of such objects related to the absence of a horizon,
which makes it possible to observe sources of radiation at any
point of the tunnel, if the tunnel medium is transparent. A
certain behavior is expected for the variations of the spectrum,
flux, and polarization of the source of radiating objects making
periodic oscillations relative to a WH throat could be detected.

The observer structures of the sources that could be related To
WHs entrances (or BHs) with a radial magnetic field could be
studied with microarcsecond or better angular resolution, as is
proposed in the "Radioastron"\, and "Millimetron"\, space-VLBI
experiments \cite{38}, \cite{39}.

It may be possible to detect sources associated with binary WHs
entrances, which form systems with strong magnetic-dipole
radiation and the ejection of relativistic ${e^\pm}$. The final
stage in the evolution of such systems is the formation of a BH,
accompanied by strong electromagnetic impulses.

Note also that the possible existence of WHs with strong magnetic
fields suggests that theoretically predicted elementary magnetic
monopoles \cite{38}, \cite{39} could have merged with these
objects in the course of their cosmological evolution.

Another direction for future studies is related to spectral and
polarization monitoring of these sources.

The detection of tunnels will open the way to studies of the
entire Multiverse.

In the Appendixes we consider the explicit mathematical models of
the WHs with magnetic field which are the theoretical basis of our
hypothesis.

\section{ACKNOWLEDGEMENTS}
\label{sa} This work was supported by the Russian Foundation for
Basic Research (project numbers: ${05-02-17377}$,
${05-02-16302-a}$, ${04-02-16987-a}$, ${04-02-17257-a}$) and the
Program in Support of Leading Scientific Schools of the Russian
Federation ${(NSH-1653.2003.2)}$ and program "stellar evolution".

The authors are grateful to S.P. Gavrilov~\cite{40}, B.V. Komberg,
M.B. Mensky, D.I. Novikov, V.I. Ritus, and A.E. Shabat for useful
discussions and comments.

\section{APPENDIXES}
\label{s5}
\subsection{Spherically symmetrical wormhole with radial magnetic field}
\label{a1} The metric of a spherical WH can be presented in the
form (see~\cite{15}):
\begin{equation}
ds^2=e^{2\phi(r)}\, dt^2 - {dr^2\over 1-b(r)/r} - r^2\,
d\Omega^2\, , \label{7}\end{equation} where $r$ is the radial
coordinate, ${\varphi(r)}$ the so-called redshift function, and
${b(r)}$ the form function. The WH neck corresponds to the minimum
${r = r_0 = b(r_0)}$ and ${b_{(r_0)}\le 1}$. The presence of a
horizon corresponds to the condition ${\varphi\to -\infty}$ or
${e^\varphi \to 0}$; for WH $\varphi$ must be finite everywhere.

Let us introduce the mass ${M(r)}$ of the WH for an external
observer:
\begin{equation}
M(r)=M_0 +\int\limits_{r_0}^r 4\pi\varepsilon r^2\, dr \, ,
\label{8}\end{equation} where ${M_0 = r_0/2}$ and ${\varepsilon
(r)}$ --- is the energy density. For convenience in graphical
representation and calculations, we introduce the variable ${x =
r_0/r}$. The entire interval ${r_0 \le r < \infty}$ will then be
transformed into ${0 < x \le 1}$, and we obtain the equation for a
static WH:
\begin{equation}
\begin{array}{lll}
8\pi\varepsilon r_0^2= -b' x^4/r_0\, , \\
\quad \\
8\pi p_\parallel r_0^2=-bx^3/r_0 -2x^3 (1-bx/r_0) \phi' \, , \\
\quad \\
8\pi p_\perp r_0^2= (1-bx/r_0)\left[ x^4\phi'' +x^3\phi'
+x^4(\phi')^2 \right] +0.5 x^3 (xb'+b) (1 - x\phi')/r_0 \,  .
\end{array}
\label{9}\end{equation} where the derivatives are taken with
respect to $x$. It was shown in \cite{16} that, in a spherically
symmetrical WH with ${w_\parallel = const}$ and ${w_\perp =
const}$, an inequality specifying possible equations of state and
their anisotropy must be satisfied:
\begin{equation}
-2w_\perp < w_\parallel < -1 \label{10}\end{equation} The
left-hand side of the inequality specifies the finiteness of the
WH mass as ${r\to\infty}$, while the right-hand side indicates the
absence of a horizon.

It is of great interest that condition (\ref{10}) is "almost"\,
satisfied for a magnetic (or electric) field. If the field is in
the $r$ direction, the energy-momentum tensor specifies the
equation of state
\begin{equation}
w_\parallel = -1\, ,\quad w_\perp = 1\, ,\quad \varepsilon = (H^2
+ E^2)/(8\pi)\, , \label{11}\end{equation} which satisfies the
conditions (\ref{10}) for a WH to within a small negative
increment in ${w_\parallel}$.

In \cite{17}, a model for a phantom-matter WH with an anisotropic
equation of state is considered:
\begin{equation}
1 + \delta = -p_\parallel /\varepsilon = p_\perp /\varepsilon \, ,
\label{12}\end{equation} and it is shown that even an arbitrarily
small $\delta$ is sufficient for a WH to exist.

Let us denote ${x_h = r_0 /r_h > 1}$ to be the ratio of the radii
of the WH throat and the horizon of the corresponding
Reissner-Nordstrem black hole (BH)~\cite{18} with magnetic charge
${Q = r_h}$ (see~\cite{our}). $\varepsilon$ is specified by the
relation
\begin{equation}
\varepsilon =\varepsilon_0\, x^4\, \left[ (x_h-1)/(x_h-x)
\right]^{\delta}\, , \quad \varepsilon_0 = 1/(8\pi
r_0^2(1+\delta))\, . \label{13}\end{equation} For an observer far
from the throat, the WH mass ${M_\infty}$ is in the interval
\begin{equation}
M_0 \le M_\infty \le 2M_0 \label{14}\end{equation} The left-hand
side of the inequality follows from (\ref{8}), and the right-hand
side from (\ref{13}).

In this connection, it may be concluded that the electromagnetic
(EM) field may be an appreciable or even predominant part of the
WH. Let us consider three types of models:

(1) an EM field is the main component of the WH matter. In
addition a small amount of the phantom energy is necessary to
provide conditions above,

(2) an EM field plus phantom energy with an isotropic equation of
state,

(3) an EM field plus phantom matter with an anisotropic
(vector-type) equation of state.

Common to all three models is the assumption that the WH is
penetrated by an initial magnetic field, which should display a
radial structure for an external observer in the spherically
symmetrical case; i.e., it should be correspond to a macroscopic
magnetic monopole (${r\to\infty}$, ${H \propto r^{-2}}$, and
${\varepsilon_{_H} \propto r^{-4}}$). The two entrances to the WH
should have opposite signs of the magnetic field.

In model 1 there is the magnetic field plus a small amount of
phantom energy or phantom matter. Under real conditions, this
model is specified by a strong magnetic field. The fact that
$\delta$ is small indicates that the configuration is close to a
BH. The accretion of normal matter onto the WH entrance results in
the growth of ${w_\parallel}$, and the condition of the absence of
a horizon can be violated (the right-hand side of (\ref{10})). The
WH entrance can then turn into a BH with the radial magnetic
field. The opposite is also true: the accretion of phantom energy
makes the WH more different from a BH \cite{19}. Overall, model 1
is close to a BH with an extremely strong magnetic field
\cite{20}, which, however, displays a monopole rather than a
dipole structure.

TABLE 1 presents the parameters for the throats of magnetic WHs
with various masses ${M_0}$. We can use (\ref{13}) (and the
constants $c$ and $G$) to derive expressions for $r_0$, $H_0$,
$\rho_0$ (the mass density), ${\nu_{_G}}$ (the frequency of
oscillations with a small amplitude for a sample particle), and
${\nu_{_H}}$ (the gyrofrequency) in the throat and its rest frame;
$\nu_c$ is the frequency of revolution along the lower stable
circular orbit for an external observer. Then,
\begin{equation}
\begin{array}{l}
r_0=(G/c^2)\cdot M_\infty \, ;\\
H_0=(c^4) /(G^{3/2})\cdot M_\infty^{-1}\, ;\\
\rho_0=(c^6)/(8\pi G^3)\cdot  M_\infty^{-2}\, ;\\
\nu_{_G}=(c^3)/(2\sqrt{2}\pi G)\cdot M_\infty^{-1}\, ;\\
\nu_{_H}=(e c^3)/(2\pi m_e G^{3/2})\cdot M_\infty^{-1}\, ;\\
\nu_c=(\sqrt{3}c^3)/(32\pi G)\cdot
M_\infty^{-1}=\sqrt{(3/128)}\cdot \nu_{_G}\, .
\end{array}
\label{15}\end{equation}

\begin{center}
TABLE 1\\
Parameters of the throats of magnetic WHs with various masses${{}^*}$:\\
\begin{tabular}{|c||c|c|c|c|c|c|}
\hline $M_\infty=2M_0$ & $r_0$, cm & $H_0$, Gs &
$\rho (r_0)$, g/cm${}^3$ & $\nu_{_G}$, Hz & $\nu_{_H}$, Hz & $\nu_c$, Hz \\
\hline\hline $6\cdot 10^{42}$ g$=3\cdot 10^{9}M_\odot$ & $4.5\cdot
10^{14}$ & $7.8\cdot 10^{9}$ &
$2.7\cdot 10^{-3}$ & $7.6\cdot 10^{-6}$ & $2.2\cdot 10^{16}$ & $1.16\cdot 10^{-6}$\\
(Quasar) & & & & (1.5 Days) & & ($9.8$ Days)\\
\hline $10^{39}$ g$ = 5\cdot 10^5M_\odot$  &  $7.4\cdot 10^{10}$ &
$4.4\cdot 10^{13}$ & $9.7\cdot 10^{4}$ & $0.045$ & $1.3\cdot 10^{20}$ & $6.9\cdot 10^{-3}$\\
($e^\pm$ pair creation) &  &  & & ($22$ s) & & ($2.4$ min)\\
\hline $2\cdot 10^{33}$ g$=M_\odot$ &  $1.5\cdot 10^{5}$ &
$2.3\cdot 10^{19}$ &
$2.4\cdot 10^{16}$ & $2.3\cdot 10^4$ & $6.6\cdot 10^{25}$ & $3.5\cdot 10^{3}$\\
(Sun) &  & & & & &  \\
\hline $6\cdot 10^{27}$ g$=M_\oplus$ & $0.45$ & $7.8\cdot 10^{24}$
&
$2.7\cdot 10^{27}$ & $7.6\cdot 10^9$ & $2.2\cdot 10^{31}$ & $1.16\cdot 10^{9}$\\
(Earth) &  & & & &  & \\
\hline $5\cdot 10^{10}$ g &  $3.5\cdot 10^{-18}$ & $10^{42}$ &
$4.4\cdot 10^{61}$ &
$9.7\cdot 10^{26}$ & $2.7\cdot 10^{48}$ & $1.5\cdot 10^{26}$\\
(positronium) & & & & & & \\
\hline $1.8\cdot 10^3$ g & $1.3\cdot 10{-25}$ & $2.6\cdot 10^{49}$
&
$3\cdot 10^{76}$ & $2.6\cdot 10^{34}$ & $7.3\cdot 10^{55}$ & $4\cdot 10^{33}$\\
($\mu^\pm$ pair creation) & & & & & & \\
\hline $2\cdot 10^{-5}$ g & $1.5\cdot 10^{-33}$ & $2.3\cdot
10^{57}$ &
$2.4\cdot 10^{92}$ & $2.3\cdot 10^{42}$ & $6.6\cdot 10^{63}$ & $3.5\cdot 10^{41}$\\
(Planck mass) & & & & & & \\
\hline
\end{tabular}
\quad\\
\quad\\
Binary WH entrances${{}^*}$:\\
\begin{tabular}{|c||c|c|c|c|}
\hline ${M_\infty}$ &
${I_{mag}}$, erg/s & ${d_{em}}$, cm & ${T_{em}}$, s & ${V_{em}}$, km/s\\
\hline\hline $6\cdot 10^{42}$ g$\approx 3\cdot 10^{9}M_\odot$ &
$7.0\cdot 10^{40}$ & $3.0\cdot 10^{19}$ & $8.2\cdot 10^{11}$ & ${1100}$\\
(quasar) & & & & \\
\hline $3.2\cdot 10^{35}$ g$\approx 160\cdot M_\odot$
& $4.0\cdot 10^{31}$ & $4.2\cdot 10^{14}$ & $1.9\cdot 10^{8}$ & ${70}$\\
($e^\pm$ pair creation) &  & & & \\
\hline $2\cdot 10^{33}$ g$=M_\odot$ &
$4.6\cdot 10^{28}$ & $1.4\cdot 10^{13}$ & $1.5\cdot 10^{7}$ & ${30}$\\
(Sun) & & & & \\
\hline $9\cdot 10^{18}$ g ${{}^{**}}$ & $3.4\cdot 10^{9}$ &
$3.8\cdot 10^{3}$ & $1$ &
${120}$ m/s\\
(Micropulsar) & & & & \\
\hline $1.8\cdot 10^{3}$ g & $4.0\cdot 10^{-12}$ & $1.3\cdot
10^{-7}$ & $1.4\cdot 10^{-8}$ &
${30}$ cm/s\\
($\mu^\pm$ pair creation) & & & & \\
\hline $2.2\cdot 10^{-5}$ g & $1.1\cdot 10^{-22}$ & $7.0\cdot
10^{-13}$ & $1.6\cdot 10^{-12}$ &
${1.4}$ cm/s\\
(Planck mass) & & & & \\
\hline
\end{tabular}
\, \\
${{}^{*}}$See appendix~\ref{a1}.\\
${{}^{**}}$Binary WH with orbital period 1 s.
\end{center}

In TABLE 1, WH parameters are estimated for quasar cores, as well
as for objects with the field (and corresponding WH mass) critical
for the creation of electron-positron pairs, with masses of the
order of the Sun's and the Earth's, with the magnetic field (and
WH mass) critical for stability of the positronium atom, with the
magnetic field (and WH mass) critical for the creation of
monopole-anti-monopole pairs, and with the Planck mass.

The magnetic field for a model described by (\ref{13}) with a
small $\delta$ will be
\begin{equation}
H \approx M_\infty\sqrt{G}/r^2 \label{16}\end{equation} Assuming
that the electric field is small, the constraint associated with
the creation of electron-positron pairs is removed (although we
present it in the TABLE 1 for the limiting case). This field, ${H
= m^2_e c^3/\mu_{_H} \approx 4.4\cdot 10^{13}\, G}$, specifies
specific conditions related to the fact that, at large fields, the
Landau excitation level exceeds the electron rest energy. The
positronium atom becomes stable in fields above ${10^{42}\, G}$,
and the medium becomes filled with such atoms created from the
vacuum \cite{21}, \cite{22}. In magnetic fields stronger than the
critical value ${H_{max}}$, a vacuum puncture and the creation of
monopole pairs occur \cite{23}-\cite{26}. If the mass of a stable,
colorless monopole \cite{27}-\cite{29} is ${m_\mu \sim 10^{16}\,
GeV \sim 10^{-8} g}$ and the magnetic charge ${\mu = (3/2)\hbar
c/e \sim 10^{-7}}$, then ${H_{max} = m^2_\mu c^3/\mu\hbar \approx
2.6\cdot 10^{49}\, G}$, and, according to (\ref{14}), the maximum
mass of a magnetic WH with this field in its throat is ${M_\infty
\approx 1.8\, kg}$. The created monopoles will be ejected from the
WH, decreasing its mass. It is not obvious whether such small WHs
are stable against other quantum processes, and the minimum mass
of the WH may turn out to be substantially higher than $1.8\, kg$.
The lower limit for the mass of a WH with a composite constitution
is obviously even lower. Note also that the absence of a horizon
for the WH results in the absence of evaporation (Hawking
radiation). Therefore, primordial low-mass WHs could be preserved
until the current epoch, unlike primordial BHs (for which the
lower limit for the mass is ${\sim 10^{15} g}$).

Model 2 corresponds to the case of phantom energy, similar to that
used in cosmological models. Since the equation of state is
isotropic, phantom energy alone cannot satisfy the left-hand side
of (\ref{10}), and a larger fraction of magnetic field is needed.
Let us denote the fraction of the magnetic-energy density
${\eta_{_H} = \varepsilon_{_H}/\varepsilon}$ and the fraction of
the phantom-energy density ${\eta_w = \varepsilon_w
/\varepsilon}$, ${\varepsilon = \varepsilon_{_H} +
\varepsilon_w}$. If the pressure is ${p_w = w\varepsilon_w}$ and
${\eta_{_H} + \eta_w = 1}$ for the phantom matter, then
${p_\parallel = -\varepsilon_{_H} + w\varepsilon_w}$, ${p_\perp =
\varepsilon_{_H} + w\varepsilon_w}$, and it follows from
(\ref{10}) with conditions ${w_\parallel = const}$ and ${w_\perp =
const}$ that
\begin{equation}
3w/(3w-1)<\eta_{_H}<1\, , \quad 0<\eta_w <1/(1-3w)
\label{17}\end{equation} As ${w \to -1}$, the fractions of the
magnetic and phantom energies are ${0.75 < \eta_{_H} < 1}$ and ${0
< \eta_w < 0.25}$. As ${w\to -\infty}$, ${\eta_{_H}\to 1}$, and
${\eta_w \to 0}$. Thus, for an electromagnetic WH, it is necessary
and sufficient to include an arbitrarily small fraction of phantom
energy with an isotropic equation of state, with w arbitrarily
close, but smaller than ${-1}$.

\begin{figure}[t]
\centering \epsfbox[100 250 550 550]{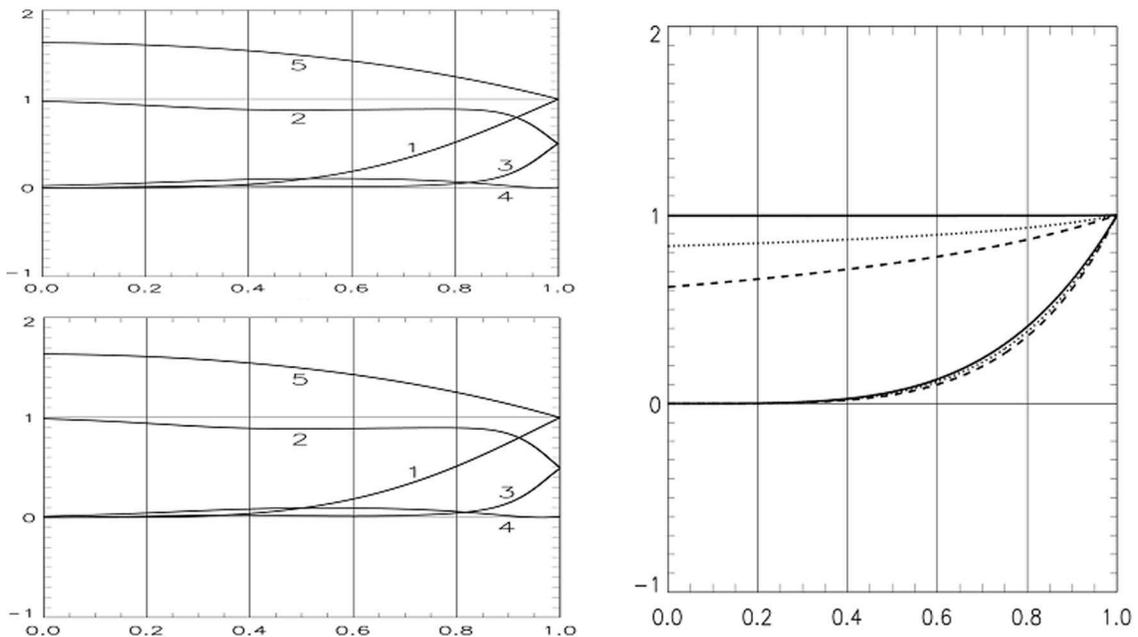} \caption{Basic model
parameters as functions WH of ${x = r_0/r}$. The top-left plots
shows the three-component model with ${w_{_H} = 1 ,\,\, w = -1}$,
and ${\eta_{_H}(1) = \eta_w(1) = 0.5}$. The dependences are
indicated for (1) the total density in a throat, (2) the
magnetic-field density from the total density, (3) the
phantom-energy density, (4) the ordinary-matter density, and (5)
the total density in a throat not divided by ${x^4}$ (this
characterizes the difference of distribution for the density from
that for a magnetic monopole). The bottom-left plots show the same
dependences for the three-component model for ${w_{_H} = 1 ,\,\, w
= 1.02}$, and ${\eta_{_H}(1) = \eta_w(1) = 0.5}$. The
right-hand-side plots show the model for a WH supported mainly by
a magnetic field (or by phantom matter with the same $\delta$).
The three lower curves represent the dependence of the fractional
density on the density in the throat ${\eta =\varepsilon
/\varepsilon_0}$ for ${\delta = 0.5}$ (dashed), ${\delta = 0.1}$
(dotted), and ${\delta = 0.001}$ (solid); the three upper curves
correspond to the lower curves with ${\eta /x^4}$ (characterizing
the difference of the structure from a monopole). As ${\delta\to 0
\,\, \eta /x^4 \to 1}$. About notations see the end of section
\ref{a1}.} \label{R2}
\end{figure}

Figure~\ref{R2} presents the results of calculating equations
(\ref{9}) for all three models. It shows the dependence of the
fractions of ${\eta_{_H}}$, ${\eta_w}$, and ${\eta_{_{CDM}}}$ on
${x = r_0/r}$ for the models with a radial magnetic field, phantom
energy (with ${w \to -1}$), and ordinary matter with "zero"\,
pressure (the fraction ${\eta_{_{CDM}}}$). The limiting conditions
are ${\eta_{_H} = \eta_w =0.5}$ in the throat, ${\eta_w \to
\eta_{_{CDM}} \to 0}$  for ${x\to 0\,\,(r\to \infty)}$. Here, in
the three-component model, unlike (\ref{12}), ${w_\parallel}$ and
${w_\perp}$ depend on $r$.

Model 3 can provide any relation between the magnetic and phantom
matter density, while, according to~\cite{17}, $\delta$ can be
arbitrarily small.

We want to emphasize that the relations between $p$ and
$\varepsilon$ correspond to the static state and we do not
consider the static state and we do not consider the dynamics.
Particularly we do not consider the problem of stability of
wormholes.

\subsection{DIPOLAR ELECTRIC FIELD INDUCED IN A WH}
\label{a2} In the case of a rotating WH with a magnetic field, the
electric field that is induced should display a dipolar structure,
due to the monopolar WH magnetic field.

Characteristic estimates for the induced WH electric field can be
obtained from the exact solution corresponding to a BH~\cite{41}.
Making the substitution "electricity $\longleftrightarrow$
magnetism"\, in the solution, we obtain for the electric field:
\begin{equation}
\vec E(r,\theta) \approx H(r_h)\cdot {a\cdot r_h^2\over r^3}\cdot
\left(2\, \cos\theta \cdot\vec e_r +\sin\theta\cdot\vec e_\theta
\right) \label{18}\end{equation} Observationally, the existence of
such a field in the vacuum approximation in a toy model will
result in the ejection of electrons from one of the poles and
protons from the other. Standard Blandford-Znajek type~\cite{42}
hydrodynamical models or models with a quadrupolar electric field
predict two-sided jets for BHs. In addition, even in the case of
slow rotation, such a jet should display a higher energy, since
its acceleration results from a larger electric potential then in
the case of a BH: ${\varphi_e \sim H(r_0)a}$ for a WH, while
${\varphi_e \sim H(r_h)a^2/r_h}$ for a BH, where $a$ is the
rotation parameter in the Kerr metric.

\subsection{OBSERVATIONS OF BODY OSCILLATING THROUGH A WH THROAT}
\label{a3} Oscillations of bodies in the vicinity of a WH throat
(radial orbits) could give rise to a peculiar observational
phenomenon. Signals from such sources detected by an external
observer will display a characteristic periodicity in their
spectra. All objects (stars, BHs) other than WHs absorb bodies
falling onto them irrecoverably. Periodic radial oscillations are
a characteristic feature of WHs.

For simplicity, let us consider a test body with zero angular
momentum. We will solve the equations of motion using the
Hamilton-Jacobi method in a curved space~\cite{33}.

We then obtain for the velocity of the body:
\begin{equation} {\partial r\over \partial t}=\pm\exp (\phi)\cdot \sqrt{\left[ 1-b/r \right]\cdot \left[1-\exp(2\phi) \cdot (m_0/E_0)^2\right] } \label{19}\end{equation} Here, ${m_0}$ and ${E_0}$ are the rest mass and total energy
of the body.

Expression (\ref{19}) does not correspond to the physical velocity
of the body, since $r$ is not the physical radial coordinate. The
physical velocity of the body along the radius, ${\dot l}$ (with
respect to time $t$) is
\begin{equation} \dot l = \pm\exp (\phi)\cdot \sqrt{1-\exp(2\phi) \cdot (m_0/E_0)^2 } \label{20}\end{equation} The redshift of a signal radiated by the body is
specified by the following two factors.

(1) The Doppler shift due to the motion of the source yields the
factor ${\sqrt{1 - v^2}/(1 \pm v)}$, where ${v = \dot
r\sqrt{|g_{rr}/g_{tt}|}}$ is the physical velocity of the body in
its proper time (here, the signs $\pm$ correspond to motion of the
body from and toward the observer).

(2) The gravitational redshift yields the factor ${\exp\varphi}$.

Thus, the frequency of the signal received by a distant observer
will be given by the expression
\begin{equation} \nu = \nu_0\cdot  {\exp(2\phi)\cdot (m_0/E_0) \over
1\pm\sqrt{1-\exp(2\phi)\cdot (m_0/E_0)^2}}
\label{21}\end{equation} Here, ${\nu_0}$ is the frequency of the
signal measured on the moving body. It follows that, unlike a BH
horizon, the frequency of the signal in the WH throat does not
become equal to zero for a distant observer.

\begin{figure}[t]
\centering \epsfbox[100 260 530 650]{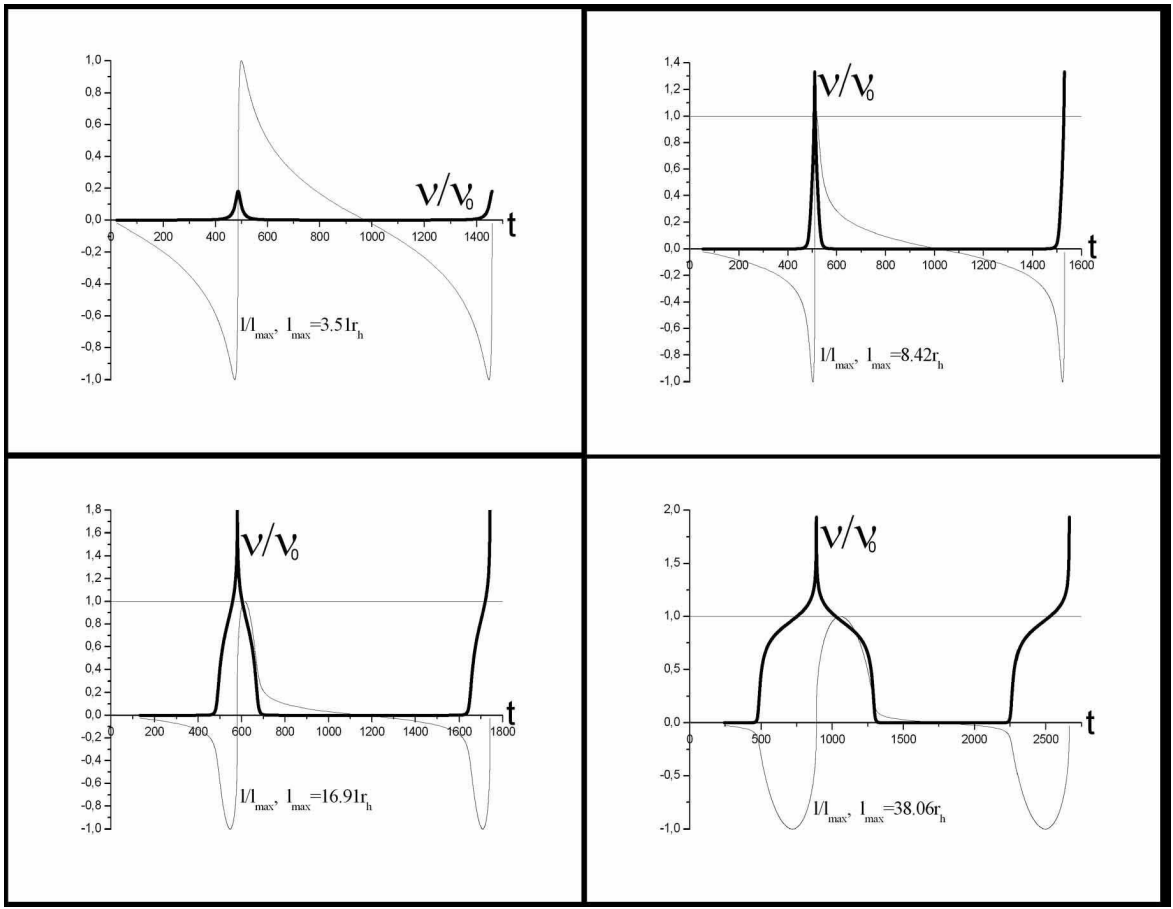} \caption{Time
dependences of the frequency shift ${\nu /\nu_0}$ (thick curve)
and physical radial coordinate ${l/l_{max}}$ (thin curve); time
$t$ is in units of ${r_h/c}$. The graphs are plotted for ${\delta
= 0.001}$.} \label{R5}
\end{figure}

In order to find the time dependence of the redshift ${\nu
/\nu_0}$ of the body for an external observer, time ${\Delta t}$
for the light to propagate from point $r$ to point ${r_{max}}$ of
the body must be added to time $t$. Figure~\ref{R5} presents these
dependencies.

For extremely small amplitudes ${(r_1 - r_0)}$, the test-body
oscillations become harmonic. This situation is reached when the
following inequality is satisfied:
\begin{equation}
r - r_0 \le r_1-r_0 << r_0 - r_h  \,  , \qquad 1-x \le 1-x_1 <<
x_h -1\, , \label{21}\end{equation} where $x_1=r_0/r_1$. In this
case, in the vicinity of the points where the body stops, the
components of its velocity ${\dot r}$ (\ref{19}) can be expanded
in a series in ${1 - x}$ (or ${x - x_1}$). Restricting the
expansion to the main terms yields
\begin{equation}
e^\phi \approx (x_h -1)\, ,\quad (1-{b\over r}) \approx (x_h
-1)(1-x)\, ,\quad 1-e^{2\phi}(m_0/E_0)^2 \approx {2(x-x_1)\over
x_h -1}\, . \label{22}\end{equation} Hence,
\begin{equation}
(\dot r)^2 \approx 2(x_h -1)^2 (1-x)(x-x_1) \, ,\quad r_h\ddot r
=\dot r {\partial \dot r\over \partial x} ={1\over 2} {\partial
(\dot r)^2\over \partial x} \, ,\quad (1-x)\, \ddot{}
=-\omega_0^2(1-x)\, , \label{23}\end{equation} where ${w_0 =
\sqrt{2}(x_h - 1)/r_h}$. This is the equation for harmonic
oscillations with the body stopping at ${x = 1}$ and ${x = x_1}$;
therefore, the period of these oscillations measured by an
external observer will be
\begin{equation}
T_1= {\sqrt{2}\pi r_h\over c (x_h -1)} . \label{24}\end{equation}
In this case, the oscillations of the physical coordinate $l$ are
also harmonic, as follows from (\ref{20}) and (\ref{22}). The body
oscillates from ${-l_1}$ to ${+l_1}$; the condition that ${l_1}$
be small, corresponding to (\ref{19}), is ${l_1 << r_h}$ (thus, in
these coordinates, the amplitude does not have to be extremely
small).

The oscillations of coordinate l have a period that is twice this
value ${(T_2 = 2T_1)}$.

\subsection{CIRCULAR ORBIT AROUND A WH}
\label{a4} The difference of the metric of the limiting
Reisner-Nordstrem and our model of the WH is negligible at
${r=2r_h}$ (or more). Thus all conclusions about circular orbit
around a WH are the same as in the limiting Reisner-Nordstrem
geometry at the corresponding distances.

We obtain (see~\cite{43},\cite{44}) for the period at the circular
orbit with respect to time $t$:
\begin{equation}
\tau =  \int\limits_0^{2\pi}{d\varphi \over \dot\varphi}={2\pi r^2
E_0/L\over c^2 (1-r_h/r)^2}\, , \label{25}\end{equation} where
${E_0}$ --- is the energy and $L$ --- is the angular momentum.

Hence, the periods of the last stable and unstable circular orbits
(according to a distant observer) are
\begin{equation}
\tau (4r_h) =  {32\pi r_h \over \sqrt{3}c}\, , \quad \tau (2r_h) =
{8\pi r_h \over  c}  \, . \label{26}\end{equation}

$$ $$
\hrule
$$ $$

\end{document}